\documentclass{aastex63}

\usepackage{hyperref,epsfig,graphicx,lineno,longtable, amssymb,multirow}

\received{***}
\revised{***}
\accepted{***}

\shorttitle{M dwarf CME}
\shortauthors{Sun et al.}
\graphicspath{{./}{figures/}}
 
\begin{document}

\title{Intense but Harmless: Exo-Space Weather Around an M Dwarf with a Single-Hemisphere Dynamo}

\author[0000-0001-5657-7587]{Zheng Sun}
\affiliation{School of Earth and Space Sciences, Peking University, Beijing 100871, China; \url{huitian@pku.edu.cn}}
\affiliation{State Key Laboratory of Solar Activity and Space Weather, National Space Science Center, Chinese Academy of Sciences, Beijing 100190, China}
\affiliation{Leibniz Institute for Astrophysics Potsdam, An der Sternwarte 16, Potsdam 14482, Germany}

\author[0000-0001-5052-3473]{Julián D. Alvarado-Gómez}
\affiliation{Leibniz Institute for Astrophysics Potsdam, An der Sternwarte 16, Potsdam 14482, Germany}

\author[0000-0002-1369-1758]{Hui Tian}
\affiliation{School of Earth and Space Sciences, Peking University, Beijing 100871, China; \url{huitian@pku.edu.cn}}
\affiliation{State Key Laboratory of Solar Activity and Space Weather, National Space Science Center, Chinese Academy of Sciences, Beijing 100190, China}

\author[0000-0003-3721-0215]{Ofer Cohen}
\affiliation{University of Massachusetts Lowell, Department of Physics \& Applied Physics, 600 Suffolk Street, Lowell, MA 01854, USA}
  
\author[0000-0002-0210-2276]{Jeremy J. Drake}
\affiliation{Lockheed Martin Solar and Astrophysics Laboratory, 3251 Hanover Street, Palo Alto, CA 94304, USA}

\author[0000-0003-1231-2194]{Katja Poppenhäger}
\affiliation{Leibniz Institute for Astrophysics Potsdam, An der Sternwarte 16, Potsdam 14482, Germany}
\affiliation{University of Potsdam, Institute for Physics and Astronomy, Karl-Liebknecht-Str. 24/25, Potsdam 14476, Germany}

\author[0000-0003-1220-1582]{Yue-Hong Chen}
\affiliation{School of Astronomy and Space Science, Nanjing University, Nanjing 210023, China}
\affiliation{Key Laboratory of Modern Astronomy and Astrophysics, Ministry of Education, Nanjing 210023, China}
\affiliation{Leibniz Institute for Astrophysics Potsdam, An der Sternwarte 16, Potsdam 14482, Germany}

\begin{abstract}
M dwarfs are among the most promising host stars in the search for habitable exoplanets. However, their active atmospheres drive intense magnetic activity,  including energetic flares and possibly coronal mass ejections (CMEs), which may pose serious threats to planetary habitability. In this study, we perform three-dimensional magnetohydrodynamic (MHD) simulations of CMEs on a fully convective M dwarf with a rotation period of 30 days, corresponding to the moderate-rotation regime. The magnetic topology driving our simulations is adopted from an exploratory global dynamo simulation of a fully convective low-mass star exhibiting a single-hemisphere magnetic configuration, which is not yet observationally confirmed. The large-scale magnetic field is mostly restricted to a single hemisphere and characterized by high-latitude polarity inversion lines (PILs), with the implication that most CMEs should originate from high latitudes. We find that these high-latitude CMEs propagate radially and away from the equatorial plane, producing only weak and spatially limited disturbances along the equatorial orbits of exoplanets. Moreover, low-latitude CMEs experience stronger drag within the dense and slow stellar wind near the equator, which significantly reduces both their propagation speeds and their overall impact on exoplanets. The resulting dynamic pressure enhancements on equatorial exoplanets caused by these CMEs are within two orders of magnitude above the quiescent conditions, much lower than those reported in previous M-dwarf CME simulations. These results indicate that, if such magnetic topologies indeed exist on M dwarfs, they may produce a relatively benign CME environment, which could be favorable for planetary habitability at face value.
\end{abstract}


\section{Introduction} \label{sec:intro}

Searching for potentially habitable exoplanets has become a popular topic in modern astrophysics. M dwarfs are among the most promising host stars for such exoplanets \citep{2016Natur.536..437A,2018Natur.563..365R,2024ARA&A..62..593H}. These stars offer several advantages in the context of habitable exoplanets detection. First, M dwarfs account for approximately 70\%–75\% of all the stars in the solar neighborhood \citep{2010AJ....139.2679B,2019AJ....157..216W}. Second, their low mass and luminosity enhance the detectability of orbiting exoplanets through radial velocity and transit methods \citep{2015ApJ...807...45D,2021A&ARv..29....1K}. Third, their lower temperatures bring the habitable zone much closer to the star, increasing the likelihood of detecting close-in planets located in the habitable zone \citep{2014PNAS..11112641K,2016PhR...663....1S}.

Despite these advantages, M-dwarf systems also pose unique challenges to planetary habitability, particularly in terms of stellar space weather \citep{2013A&A...557A..67V,2014MNRAS.438.1162V,2014ApJ...790...57C,2016ApJ...833L...4G,2017ApJ...843L..33G,2022ApJ...928..147A}. Because the habitable zone lies so close to M dwarfs, exoplanets are subject to intense stellar wind and high-energy radiation \citep{2016MNRAS.459.4088O,2023MNRAS.525.5168M}.  Such conditions may drive significant atmospheric escape, threatening the long-term habitability of these planets \citep{2017ApJ...837L..26D,2023MNRAS.522.1411S,2024A&A...683A.153V}.
Moreover, M dwarfs possess deeper convective envelopes than solar-type stars, and those with masses below $\sim$0.35 M$_\odot$ are fully convective, lacking a radiative zone \citep{1958ApJ...127..363L,1990sse..book.....K,1998A&A...337..403B,2024ARA&A..62..593H}. This internal structure leads to stronger magnetic fields and more intense magnetic activities in the stellar atmosphere \citep{2001ApJ...559..353M,2021A&ARv..29....1K}. Therefore, many M dwarfs are observed to exhibit higher flare rates compared to solar-type stars \citep{2005stam.book.....G,2016ApJ...829...23D,2019ApJ...873...97L,2023A&A...669A..15Y}. These frequent and intense flares could contribute to planetary atmospheric escape and affect the chemical composition  in planetary atmospheres \citep{2016ApJ...830...77V,2025ApJ...985..100D}.

On the Sun, large flares are often accompanied by CMEs, which are massive expulsions of plasma and magnetic fields into interplanetary space \citep{2011LRSP....8....1C,2012LRSP....9....3W,2025E&PP....9..171L}. The pre-eruption structure of CMEs is commonly considered to be a flux rope, typically formed along polarity inversion lines (PILs) where magnetic fields of opposite polarities meet \citep{2012A&A...539A.131K,2019LRSP...16....3T,2020RAA....20..165L,2023ApJ...953..148S,2024ScChE..67.1592L,2025ApJ...990...45S,2026ApJ..1000..205Q}.
Once initiated, the flux rope propagates mostly radially and evolves into a CME \citep{2005SSRv..121...91V,2024MNRAS.533L..25L}.
CMEs can compress planetary magnetospheres, leading to atmospheric escape, while their embedded magnetic fields might trigger geomagnetic storms \citep{2024SoPh..299...93S,2025GeoRL..5214040F,2025MNRAS.536.1089H,2025ApJ...982..194P}. The CME-driven shocks could also efficiently accelerate energetic particles, which can further impact atmospheric chemistry and stability \citep{1990ecpi.book.....J,2016NatGe...9..452A,2025ApJ...979..100K}. 

Based on solar flare-CME associations \citep{2009IAUS..257..233Y,2011SoPh..268..195A,2019SSRv..215...39L}, the expectation is that active M dwarfs may produce more energetic CMEs \citep{2015IAUGA..2258074D,2017MNRAS.472..876O}, which could strongly affect the atmospheres of orbiting exoplanets \citep{2007AsBio...7..167K,2007AsBio...7..185L}. Still, the role played by the strong large-scale magnetic field measured in these objects \citep{2008MNRAS.390..545D,2010MNRAS.407.2269M,2017A&A...605A..13K,2018MNRAS.479.4836L,2021A&ARv..29....1K} and its possible CME-suppression influence, modifying the solar flare-CME scalings, is a matter of active research \citep{2015IAUGA..2258074D,2018ApJ...862...93A,2019ApJ...884L..13A,2020ApJ...895...47A,2020ApJ...900..128L,2021ApJ...917L..29L}. 
Recent studies have also shown that CMEs from M dwarfs can induce intense Joule heating on nearby planets, potentially leading to atmospheric erosion \citep{2014ApJ...790...57C,2025A&A...700A.225E}. \citet{2022ApJ...928..147A} simulated CMEs from the active M dwarf AU~Mic and found that the resulting dynamic pressures could reach values up to $10^5$ times greater than those during extreme solar events at Earth. 
Such intense CME impacts may therefore pose a serious threat to the habitability of planets orbiting M dwarfs.

On the other hand, \citet{2020ApJ...902L...3B} developed a stellar dynamo model to simulate the interior of rotationally constrained, fully convective M dwarfs. The surface magnetograms generated from their model reveal magnetic fields largely restricted to a single hemisphere on the stellar surface. Although this magnetic topology remains both theoretically and observationally unconfirmed, it provides an interesting exploratory framework for investigating how this large-scale stellar magnetic topology may influence CME propagation and exoplanet space weather. In this magnetic field configuration, the dominant PILs are located at high latitudes. We hypothesize that, under such configurations, most CMEs should be launched from high latitudes and propagate away from the equatorial plane. As a result, exoplanets orbiting near the equatorial plane could be subject to fewer CME impacts, potentially creating a more favorable space weather environment for habitability. In this study, we test this hypothesis through three-dimensional magnetohydrodynamic (MHD) simulations. Our methodology and simulation setup are presented in Section~\ref{sec:model}, followed by the results in Section~\ref{sec:observation}. Finally, we discuss and summarize our findings in Section~\ref{sec:kk}.

\section{Methods} \label{sec:model}
\subsection{Magnetic Maps}

\citet{2020ApJ...902L...3B} developed a global stellar dynamo model for a fully convective M dwarf using the open-source Dedalus pseudospectral framework. Their simulation was initialized with characteristic Reynolds numbers of approximately 210--350 and vorticity-based Rossby numbers of about 0.27--0.42, with these ranges reflecting the radial variation of the Reynolds and Rossby numbers within the single stratified simulation. Following the results in \citet{2021A&A...651A..66K}, for a fully convective M-dwarf, these Rossby numbers correspond approximately to moderate rotation periods of 14--43 days.

\begin{figure}
    \centering
    \plotone{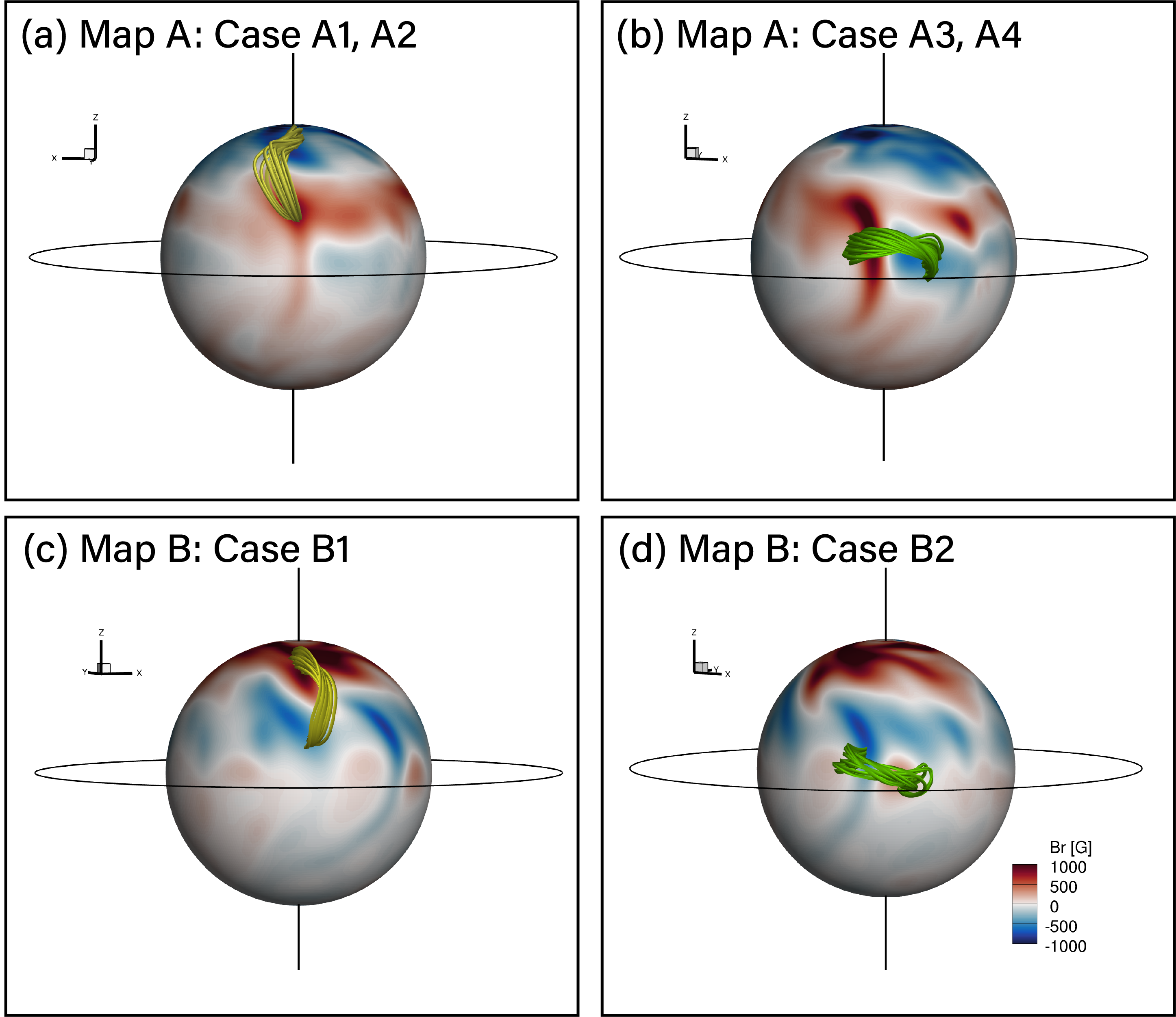}
  \caption{Magnetic maps and inserted flux ropes. Panels (a) and (b) correspond to Model A, which uses magnetic map A, while panels (c) and (d) correspond to Model B, which uses magnetic map B. The yellow and green field lines represent the inserted flux ropes, where the yellow lines indicate the high-latitude flux rope and the green lines indicate the equatorial flux rope. The black curve represents the equatorial plane.
    \label{fig:Fig.1}}
  \end{figure}

Importantly, the simulation exhibited a single-hemisphere dynamo behavior, which means most of the strong mixed-polarity magnetic fields are concentrated within one hemisphere, with cyclic polarity reversals of the large-scale magnetic field roughly every fifteen stellar rotations. This particular dynamo state remains exploratory and has not yet been observationally confirmed. Similar single-hemisphere magnetic configurations have also not been widely reported in other global dynamo simulations, possibly because the model of \citet{2020ApJ...902L...3B} adopts a distinct numerical setup that extends the computational domain all the way to the stellar center ($r=0$) without introducing an inner boundary cutout. The upper boundary of their model corresponds to the top of the convection zone, and the magnetic field there can be regarded as a quasi-representation of the photospheric magnetic configuration. Therefore, we use the magnetic map at the top of their model as the lower boundary condition for our stellar atmosphere simulation, a method that has also been adopted in previous studies (e.g. \citealt{2019ApJ...884L..13A,2020ApJ...895...47A,2024ApJ...971..153X,2025ApJ...985..219X,2025arXiv251012969C}). To represent two opposite magnetic phases, we selected two timesteps from their simulation outputs that differ in the large-scale polarity, and adopted the corresponding surface radial magnetic field ($B_r$) maps as the lower boundary conditions for our simulations. These two maps are shown in Figure \ref{fig:Fig.1}, with model A (upper row) and model B (lower row) corresponding to the two maps. Using these photospheric magnetic maps, we reconstructed the global coronal magnetic field via a potential field source surface (PFSS) extrapolation, which provides the large-scale magnetic topology used to initialize the simulations.

From the two selected magnetic maps, the PIL regions separating strong positive and negative fields are predominantly located at high latitudes (approximately between $40^\circ$ and $55^\circ$), with only a few extending close to the equator. Since CMEs are commonly initiated along PILs and tend to propagate radially (at least relatively close to the star), such a configuration implies that most CMEs from this stellar type and magnetic configuration are likely to erupt toward high latitudes, away from the equatorial plane. As a result, their impact on exoplanets in equatorial orbits is expected to be small. This serves as the basis for our working hypothesis, which we will test through numerical simulations in this study.

The dynamo model of \citet{2020ApJ...902L...3B} is dimensionless and therefore provides the large-scale magnetic geometry rather than an absolute magnetic field strength. To construct a physically scaled coronal model, we need to normalize the input magnetic field using observationally motivated values for fully convective M dwarfs with rotation periods of 14--43 days. Zeeman-Doppler imaging (ZDI) measurements of fully convective M dwarfs show large-scale mean magnetic field strengths spanning from tens of gauss to about a kilogauss. Among them, rapidly rotating (\(P_{\rm rot}<10\) days) fully convective M dwarfs can exhibit large-scale magnetic fields of \(\sim50\)--1600 G, while the limited ZDI measurements available for more slowly rotating objects (\(P_{\rm rot}>70\) days) indicate field strengths of \(\sim200\)--275 G \citep{2021A&ARv..29....1K,2021MNRAS.500.1844K}. Guided by the ZDI-inferred large-scale field strengths of fully convective M dwarfs in the rapidly rotating (\(P_{\rm rot}<10\) days) and slowly rotating (\(P_{\rm rot}>70\) days) regimes, and given the current lack of ZDI measurements for those in the intermediate rotation-period range, we adopt \(\langle |B| \rangle=300\) G at \(\ell_{\max}=5\) as a plausible representative large-scale normalization for a moderately rotating case. We then scale the original magnetic map such that its \(\ell_{\max}=5\) component satisfies this normalization, which corresponds to a surface-averaged unsigned field strength of \(\sim370\) G in the full-resolution magnetogram. We do not use the total magnetic field strengths inferred from Zeeman broadening for the normalization because Zeeman broadening measurements include contributions from both large- and small-scale magnetic fields \citep{2021A&ARv..29....1K}, whereas the dynamo solution adopted here primarily represents the large-scale magnetic field. Therefore, ZDI-inferred large-scale field strengths provide a more appropriate basis for the normalization. The \(\ell_{\max}=5\) component is adopted here only as an approximate proxy for the large-scale magnetic field typically probed by ZDI, rather than as a direct equivalent of a ZDI reconstruction \citep{2019MNRAS.483.5246L}. Accordingly, the adopted stellar and magnetic parameters should therefore be understood as representative values chosen within the expected ranges for moderately rotating fully convective M dwarfs, rather than as the parameters of a specific observed star.

\subsection{Numerical models}
We perform our simulations using the Space Weather Modeling Framework (SWMF; \citealt{2005JGRA..11012226T,2010ApJ...725.1373V,2012JCoPh.231..870T,2022ApJ...925..146V}), which is originally developed for modeling various components of the space environment of the solar system, including the solar corona, planetary magnetospheres, the outer heliosphere, etc. In recent years, the SWMF has been successfully extended to stellar environments beyond the solar system (e.g., \citealt{2011ApJ...733...67C,2011ApJ...738..166C,2020ApJ...897..101C,2014MNRAS.438.1162V,2016ApJ...833L...4G,2017ApJ...843L..33G,2022ApJ...941L...8G,2016A&A...594A..95A,2016A&A...588A..28A,2020MNRAS.494.2417V}). In this work, we primarily use the Alfvén Wave Solar atmosphere Model (AWSoM; \citealt{2014ApJ...782...81V}) module within the Solar Corona (SC) component of SWMF. AWSoM solves the three-dimensional magnetohydrodynamic (MHD) equations using the Block Adaptive-Tree Solar-wind Roe-type Upwind Scheme (BATS-R-US; \citealt{1999JCoPh.154..284P}) in spherical coordinates in SC. This model is a two-temperature model and incorporates the propagation, reflection, and dissipation of Alfvén waves, which contribute to coronal heating and stellar wind acceleration. 
The Alfvén-wave Poynting flux injected at the stellar surface is specified through a fixed solar-calibrated ratio \(S/B\), so that the wave energy input scales self-consistently with the local magnetic field strength of the adopted magnetic map.

Our stellar CME simulation consists of two main steps. In the first step, we construct a steady-state stellar wind model using AWSoM to simulate the background stellar corona and wind environment. Fully convective M dwarfs are generally expected below \(M_\star\simeq0.35\,M_\odot\), with M dwarfs spanning down to the hydrogen-burning limit near \(0.08\,M_\odot\) \citep{2015ApJ...804...64M,2021A&ARv..29....1K}. We therefore adopt representative stellar parameters of \(M_\star=0.15\,M_\odot\), \(R_\star=0.18\,R_\odot\), and \(P_{\rm rot}=30\) days, corresponding to a mid-to-late fully convective M dwarf in the regime of moderate rotation. Other parameters, for which no direct observational constraints exist, are adopted from typical solar values. Specifically, at the bottom boundary the electron density is set to $2 \times 10^{10}$ cm$^{-3}$ with a temperature of $T_e=T_p=5 \times 10^4$ K. The Alfvén wave perpendicular correlation length $L_\perp \sqrt{B}$ and Poynting flux to field strength ratio $(S_A/B)_\odot$ are set to $1.5 \times 10^5$ m $\cdot$ $T^{1/2}$ and $1.1\times10^6~\mathrm{W\,m^{-2}\,T^{-1}}$, respectively \citep{2014ApJ...782...81V}. The simulation domain extends from 1 $R_\star$ to 80 $R_\star$. The magnetic maps generated by the dynamo model use a spherical harmonic truncation at $\ell_{max}=127$, corresponding to an angular resolution of approximately 1.4$^{\circ}$. To ensure consistency, we use angular resolutions of 1.4° in both latitude and longitude, and a minimum radial grid size of approximately 0.0007 $R_\star$. The grid resolution remains constant during both the steady-state and CME simulations. At the adopted resolution, the large-scale coronal structures can be reliably reproduced. Although smaller-scale magnetic elements are not fully resolved, their impact on the overall evolution or propagation of CMEs is expected to be limited.

In the second step, we initiate CMEs by inserting magnetic flux ropes into the steady-state stellar wind background. For this purpose, we use the analytical Gibson–Low (GL) flux rope model \citep{1998ApJ...493..460G}, as implemented in the Eruptive Event (EE) module of SWMF. This model has been widely adopted for CME simulations in both solar and stellar contexts (e.g., \citealt{2004JGRA..109.2107M,2017ApJ...834..173J,2018ApJ...862...93A,2020ApJ...895...47A,2024MNRAS.533.1156S}). The choice of the flux rope energy is guided by the only existing M-dwarf CME simulation with observationally motivated eruption dynamics, namely \citet{2022ApJ...928..147A}. In that study, the flux rope parameters were adjusted to produce a CME with kinetic energy consistent with the best candidate CME event observed on AU Mic. Following a similar approach, we varied the magnetic field strength of the inserted flux rope in our Case A1 simulation and selected a value that yields CME propagation speeds broadly comparable to those reported for their Case 2 simulation, approximately 4000- 5000 $km\  s^{-1}$. For Case A1, the adopted GL flux-rope magnetic field strength is 50 G, which lies within the range of magnetic field strengths inferred from Zeeman broadening measurements of fully convective M dwarfs (e.g., \citealt{2022A&A...662A..41R}). In this study, our primary focus is on high-latitude CMEs, which are expected to dominate in this type of magnetic field environment. Our goal is to evaluate how strongly such CMEs affect the equatorial plane. For comparison, we also include low-latitude CME simulations. For magnetic Map A, we conduct four simulations: two high-latitude and two low-latitude events, each with different flux rope field strengths. For Map B, we perform two simulations, one at high latitude and one at low latitude. The detailed parameters of the flux ropes are listed in Table 1 and the locations are visualized in Figure 1. It is important to note that our inference regarding the dominance of high-latitude CMEs is based solely on the spatial distribution of PILs, rather than on a fully self-consistent model that includes the flux rope formation process.

\begin{table}
\centering
\caption{Parameters of the inserted flux ropes}

\begin{tabular}{ccccccc}
Case&Latitude [deg]&Longitude [deg]&Orientation [deg]&B$_{\rm FR}$ [G]&E$_{\rm FR}$ [$10^{34}$ erg]\\
\hline
\hline
A1&48.8&91.0&102.9&50&3.8 \\
A2&48.8&91.0&102.9&100&15.2 \\
A3&6.0&291.0&332.9&50&3.8 \\
A4&6.0&291.0&332.9&100&15.2 \\
B1&40.8&242.0&250.8&50&3.8 \\
B2&0.0&302.0&129.5&100&15.2 \\
\hline
\hline
\end{tabular}

\tablecomments{%
Cases A1–A4 are based on \textit{Map A}, and B1–B2 are based on \textit{Map B}. The orientation refers to the direction of the inserted flux rope and is associated with the orientation of the corresponding PIL. B$_{FR}$denotes the magnetic field strength parameter of the Gibson--Low (GL) flux rope. E$_{FR}$ represents the magnetic energy of the inserted flux rope and is calculated as the difference in magnetic energy before and after the insertion of the flux rope.
}
\label{tab:table1}
\end{table}

\section{Results} \label{sec:observation}

\subsection{Steady-state Stellar Wind}

\begin{figure}
    \centering
    \plotone{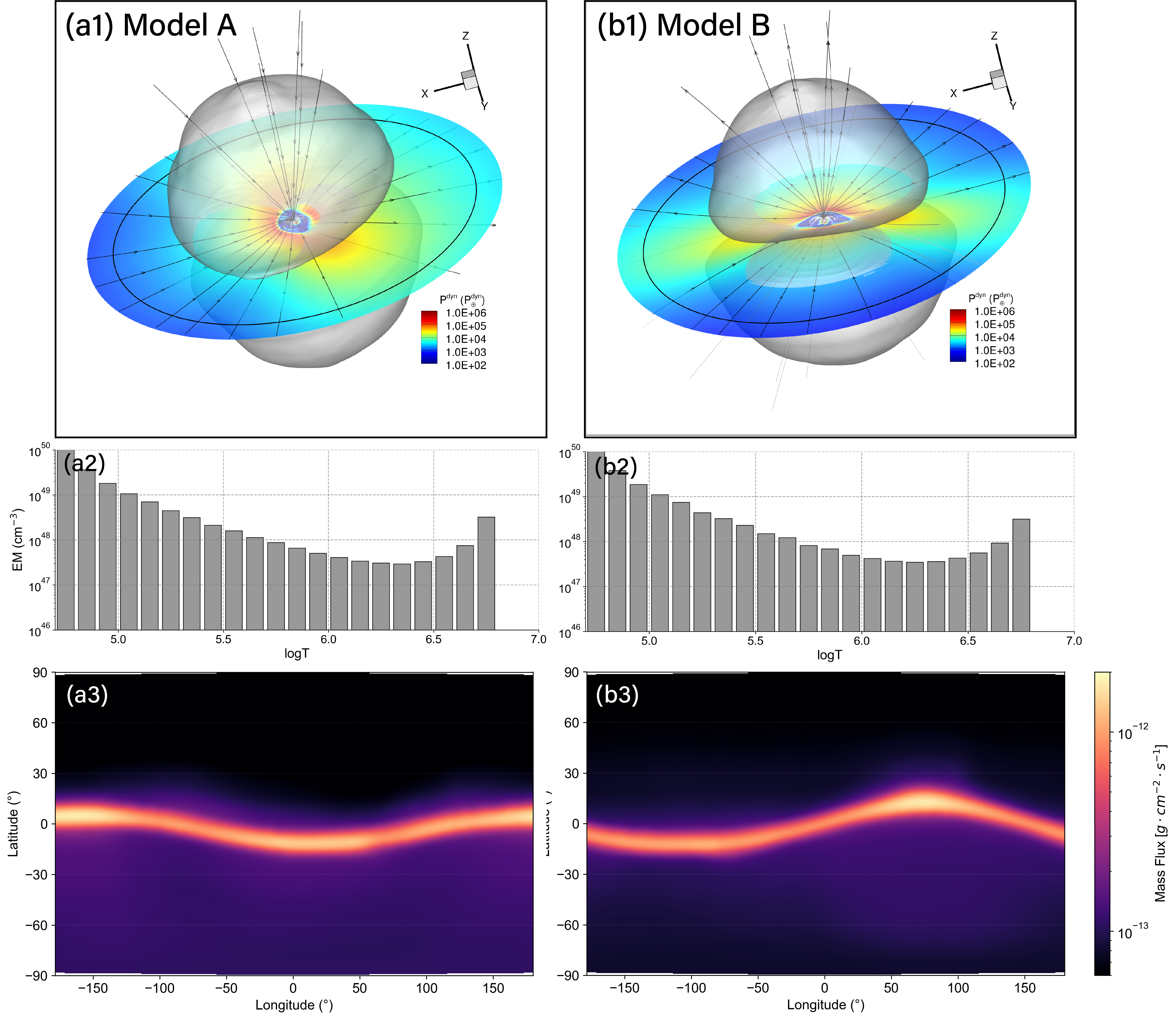}
  \caption{Steady-state stellar coronal models. Panels (a1) and (b1): Three-dimensional visualizations of Models A and B, generated using magnetic maps A and B shown in Figure \ref{fig:Fig.1}, respectively. The gray surfaces and black curves indicate the Alfvén surfaces and the hypothetical exoplanet orbit at a radius of $66\,R_\star$. Panels (a2) and (b2): Emission measure distributions of Models A and B. Panels (a3) and (b3): Mass flux distributions on the spherical surface at $R = 66\,R_\star$.
    \label{fig:Fig.2_new}}
  \end{figure}

Figures \ref{fig:Fig.2_new}(a1) and (b1) present the steady-state stellar corona generated using the two input magnetic maps, Map A and Map B, respectively. Hereafter, these two corona models are referred to as Model A and Model B. We calculated stellar mass-loss rates at 75 $R_\star$ of 
\(1.8\times10^{-14}\,M_\odot\,\mathrm{yr}^{-1}\) and 
\(1.9\times10^{-14}\,M_\odot\,\mathrm{yr}^{-1}\) from the simulations, corresponding to about \(0.9\,\dot{M}_\odot\). This value is within the range inferred for fully convective M dwarfs from Ly\(\alpha\) astrospheric absorption measurements, typically \(\sim0.2\)--\(1\,\dot{M}_\odot\) \citep{2021ApJ...915...37W}.
Additionally, we computed the emission measure distributions of both models (shown in panels (a2)--(b2)) and found that they lie between the observationally inferred emission measure distributions of slowly and rapidly rotating M dwarfs reported in \citet{Drake2020} and \citet{2025A&A...693A.285S}. It should be noted that, at lower temperatures, the comparison should be interpreted with caution because the lower corona and transition region are not fully resolved in the present model. At the stellar equator, the wind speed reaches approximately 
$\sim$400-1200 $\mathrm{km\,s^{-1}}$, which is faster than the solar and simulated solar-type stellar winds \citep{2016A&A...594A..95A,2021LRSP...18....3V}, but consistent with previous M-dwarf wind simulations \citep{2017ApJ...843L..33G,2020ApJ...902L...9A}. The gray shapes in the figures represent the Alfvén surfaces, defined as the locations where the wind speed equals the Alfvén speed. The average radial distances of the Alfvén surfaces are approximately $ 45\,R_\star $ and $ 42\,R_\star $, respectively, comparable to those obtained in previous M dwarf coronal simulations \citep{2016ApJ...833L...4G,2020ApJ...902L...9A,2022ApJ...928..147A,2023MNRAS.524.5060C}. 

Given the M4–M5 spectral type and the radius of \(0.18\,R_\odot\), the classical habitable zone estimated using the relations of \citet{2014ApJ...787L..29K} would lie between $ 66\,R_\star $ and $ 125\,R_\star $. Since this region lies entirely outside the Alfvén surface, a planet located within the habitable zone would orbit in a super-Alfvénic stellar wind environment. 
In such conditions, the relative motion between the stellar wind and the planetary magnetosphere can lead to the formation of a bow shock ahead of the planet \citep{2008ApJ...676..628S,2025ARA&A..63..299V}. 
Similar structures are observed in the solar system, for example around Earth, where the solar wind is also super-Alfvénic.
In this study, we consider a hypothetical equatorial, circular orbit at the inner edge of the habitable zone ($ 66\,R_\star $), illustrated by the black curves in Figures \ref{fig:Fig.2_new}(a) and (b). We also calculate the mass flux distribution on the spherical surface at $R = 66\,R_\star$, as shown in panels (a3) and (b3). A ribbon-like structure with enhanced mass flux density is evident near the equator, which corresponds to the equatorial streamer belt. Additionally, a clear north–south asymmetry is present in the mass flux distribution. The southern hemisphere exhibits a higher mass loss rate than the northern hemisphere, which is likely associated with the asymmetry in the input magnetic field map. Further examination reveals that this difference is primarily caused by the density distribution, with the southern hemisphere having systematically higher plasma density.

For \textit{Model A}, the stellar wind speed along the orbit ranges from 683 to 1547 $\mathrm{km\,s^{-1}}$, with the corresponding dynamic pressure $P_{\mathrm{SW}} = \rho v^2$ varying between $10^{3.1}$ and $10^{3.9}$ times the typical dynamic pressure at Earth ($P_{dyn}^{\oplus}$). 
For \textit{Model B}, the wind speed ranges from 836 to 1594 $\mathrm{km\,s^{-1}}$, while the dynamic pressure varies from $10^{3.0}$ to $10^{3.8} P_{dyn}^{\oplus}$. 
These values are consistent with those obtained in previous M dwarf wind simulations \citep{2016ApJ...833L...4G,2022ApJ...928..147A}. Moreover, the dynamic pressure varies by up to a factor of $ 10^{0.8} \approx 6.3 $ across the orbit, which may lead to substantial time-dependent effects such as enhanced Joule heating in the planetary upper atmosphere \citep{2014ApJ...790...57C,2024ApJ...962..157C}. To assess the interaction between the stellar wind and a potentially magnetized exoplanet, we compute the magnetopause standoff distance assuming pressure balance between the stellar wind and the planetary magnetic field \citep{1969JGR....74.1275S}:

\begin{equation}
\frac{R_{\mathrm{mp}}}{R_{\mathrm{planet}}} = \left( \frac{B_p^2}{4\pi P_{\mathrm{SW}}} \right)^{1/6}
\end{equation}
Adopting Earth's magnetic field strength ($ B_p \approx 0.31\,\mathrm{G} $) at the equator, the resulting magnetopause distance ranges from approximately $ 2.2\,R_\oplus $ to $ 3.0\,R_\oplus $, indicating significant compression of the planetary magnetosphere under such intense stellar wind conditions. For comparison, Earth's subsolar magnetopause typically lies at $\sim 10R_\oplus $ during quiet solar wind conditions \citep{1997JGR...102.9497S,1998JGR...10317691S}. Despite this significant reduction in magnetospheric volume, previous studies suggest that such environments may remain conducive to habitability \citep{2007AsBio...7..167K,2010Sci...327.1238T}.

\subsection{Stellar CMEs}

\begin{figure}
    \centering
    \plotone{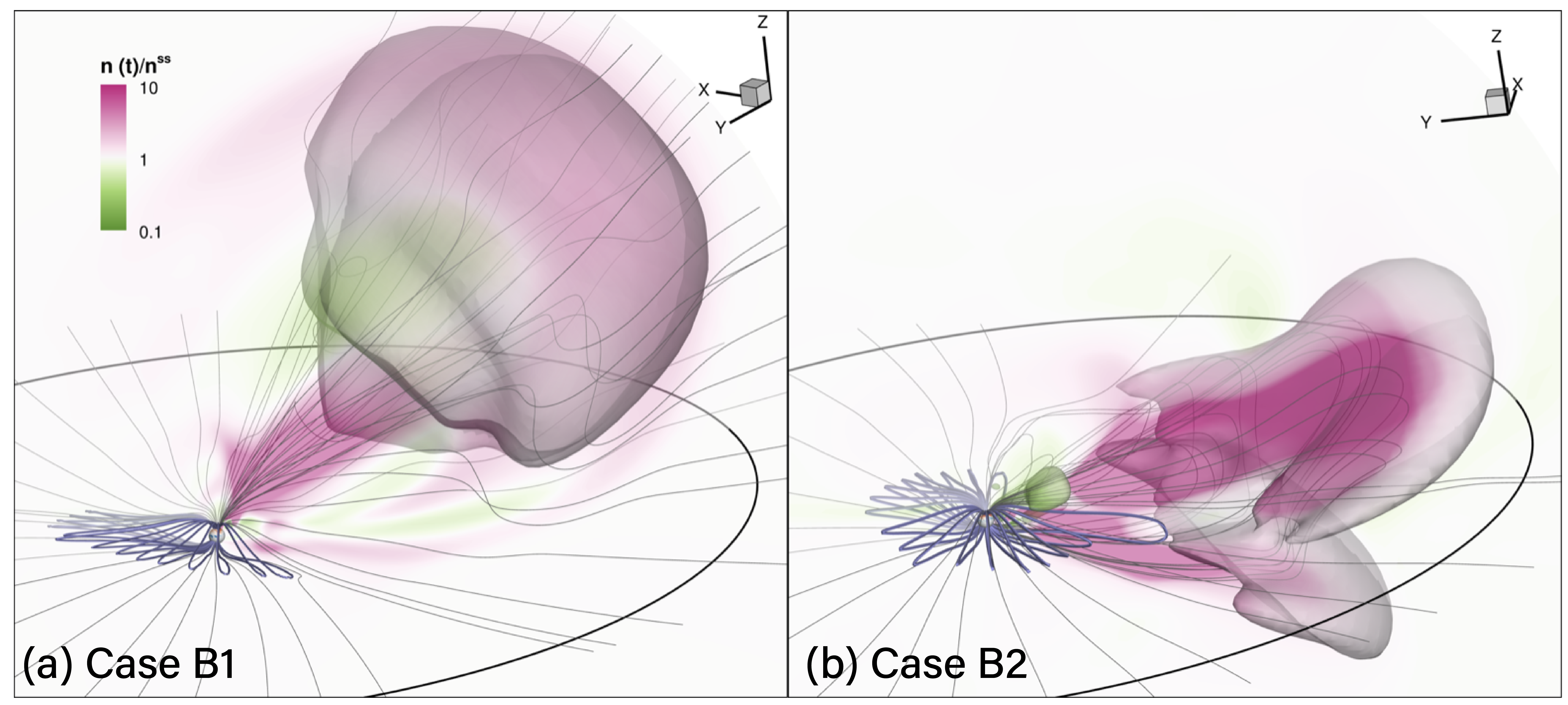}
  \caption{Snapshots of the stellar CMEs for cases B1 (a) and B2 (b). The isosurfaces represent regions where the velocity difference satisfies $u-u_{0}=3000$ and 650 km s$^{-1}$ for panels (a) and (b), respectively. The black curves indicate the hypothetical exoplanet orbit with a radius of 66 $R_\star$. (An online animation accompanies this figure. The animation shows the full temporal evolution of the CME propagation in the two cases.)
    \label{fig:Fig.2}}
  \end{figure}

The CMEs are initiated by inserting flux ropes (see Table 1 and Figure \ref{fig:Fig.1}) on the steady-state corona models. The cases A1, A2, and B1 are inserted at high latitudes, while A3, A4, and B2 are inserted near the equator. Figure \ref{fig:Fig.2} shows two representative CME results for the high-latitude case (B1) and the low-latitude case (B2). It is evident that the high-latitude one is indeed propagating toward high latitudes, away from the equatorial plane during its propagation. 
However, the low-latitude one is propagating near the equatorial plane.

\begin{figure}
    \centering
    \plotone{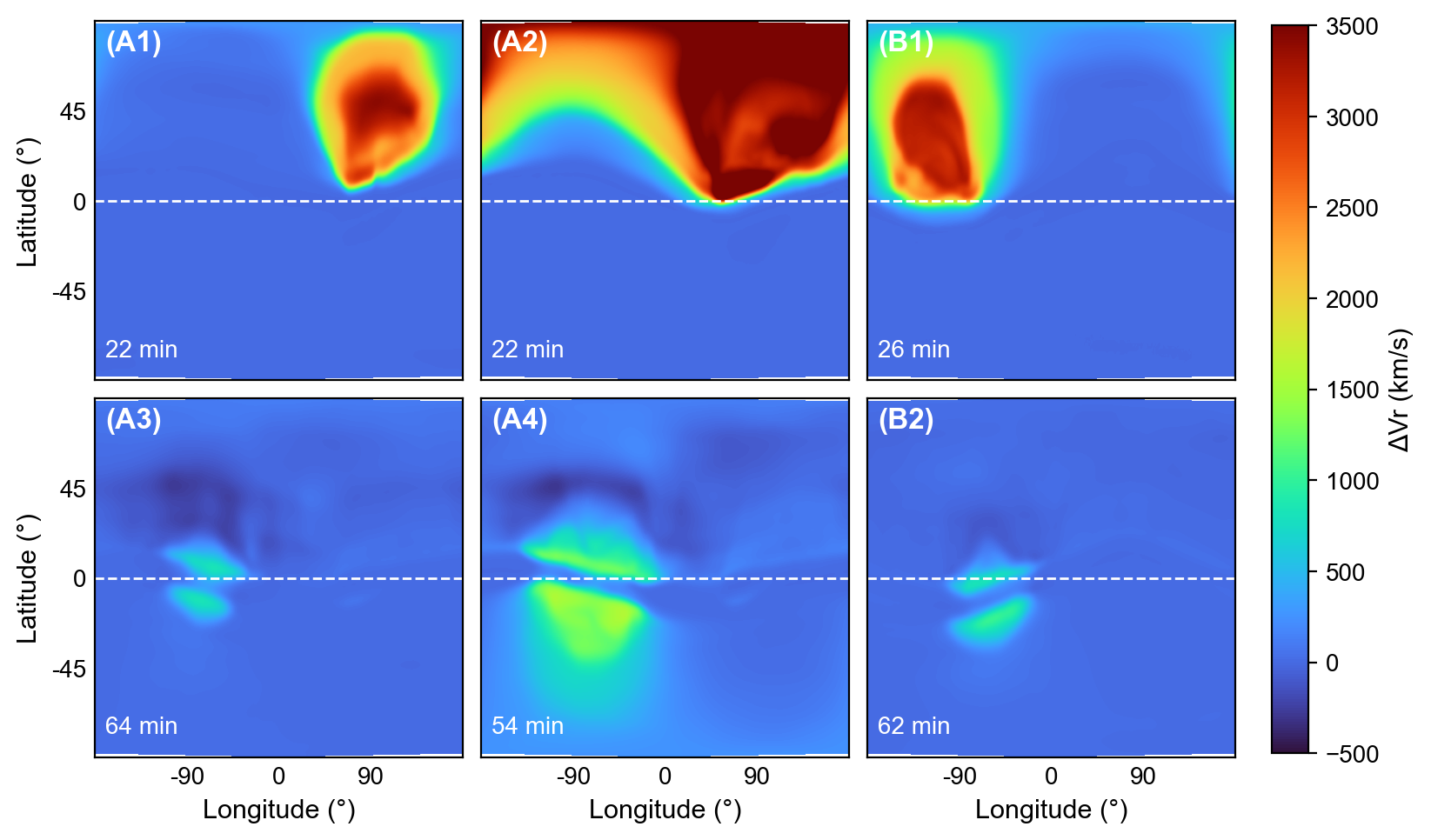}
  \caption{Velocity difference $u-u_{0}$ on a hypothetical spherical surface with a radius of $66R_\star$. The white dashed lines indicate the equatorial orbit of the exoplanet. (An online animation accompanies this figure. The animation shows the time evolution of the velocity perturbation on the spherical surface from the CME onset until the CME disturbance passes through $66R_\star$.)
    \label{fig:Fig.3}}
  \end{figure}

To quantify the spatial distribution of CME-induced velocity perturbations, we define a hypothetical spherical surface at a radial distance of $66\,R_\star$. The resulting distributions of the velocity perturbation, defined as the deviation of the instantaneous plasma velocity $u$ from the local steady-state wind velocity $u_{0}$ (i.e., $u - u_{0}$), are shown in Figure~\ref{fig:Fig.3} and the accompanying animation. We find that high-latitude CMEs (cases A1, A2, and B1) primarily disturb regions between $40^\circ$ and $60^\circ$ latitude, while the perturbations near the equator are considerably weaker. Specifically, the maximum velocity perturbations at $66\,R_\star$ reach 3707, 6921, and 5163~km~s$^{-1}$ for A1, A2, and B1, respectively, but only 2242, 3877, and 2713~km~s$^{-1}$ in the equatorial region. Notably, even for the more energetic high-latitude CME (case~A2), the equatorial velocity perturbations remain modest in both magnitude and spatial extent, as also seen in the animation. Comparing cases~A1 and~A2, we infer that increasing the flux rope energy substantially enlarges the disturbed area at high latitudes, whereas the impact on the equatorial region remains minor. For the low-latitude CMEs (cases~A3, A4, and~B2), the velocity perturbations are concentrated at low latitudes. Surprisingly, however, their magnitudes are much smaller than those of the high-latitude events, indicating that their CME speeds are significantly lower. The maximum velocity perturbations at $66\,R_\star$ on the equatorial plane are only 749, 1098, and 1342~km~s$^{-1}$, respectively. This implies that even CMEs originating near the equator will not produce substantial disturbances in the equatorial region due to their relatively low speeds. We notice a gap between the two CME segments, which corresponds to the streamer region characterized by higher plasma density and is therefore more difficult to disturb. It is also worth noting that after the passage of the CME, negative velocity perturbations appear at high latitudes (see the online animation), which may be caused by the fallback of part of the CME material \citep{2016A&A...592A..17I,2019ApJ...884L..13A,2024ApJ...974..205S}.

\begin{figure}
    \centering
    \plotone{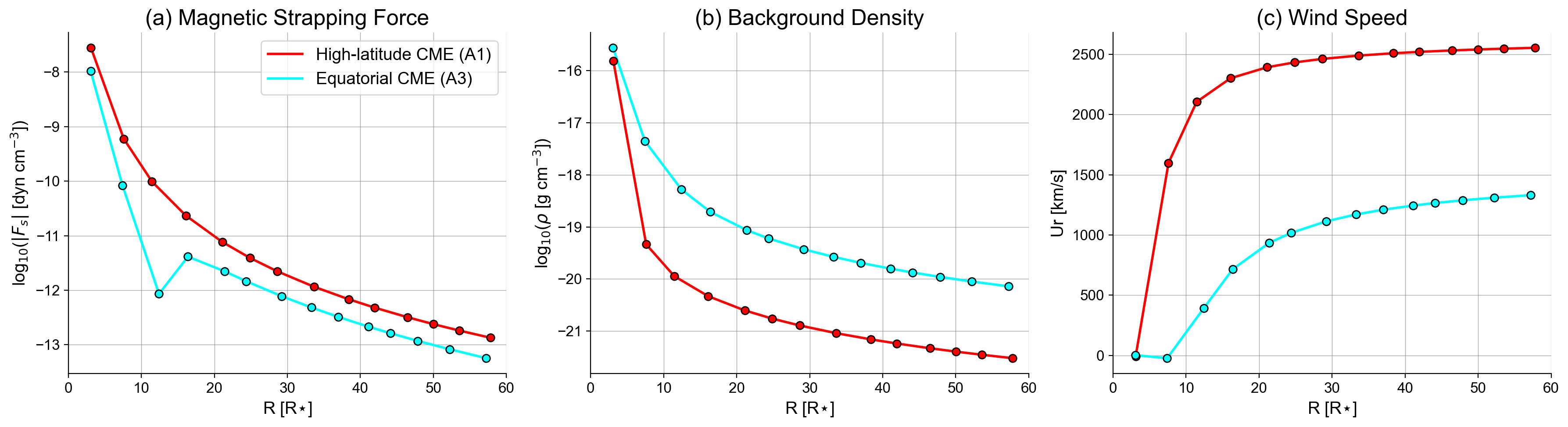}
  \caption{Average magnetic strapping force (a), background density (b), and wind speed (c) distribution along the CME propagation path calculated from the steady-state models. The red and cyan curves correspond to high-latitude case A1 and low-latitude case A3, respectively.
    \label{fig:Fig.4}}
  \end{figure}

As discussed above, the comparison between high- and low-latitude events (e.g., A1 vs.~A3, A2 vs.~A4, and B1 vs.~B2) reveals that, for the same injected magnetic energy, the equatorial CMEs propagate more slowly than their high-latitude counterparts. To investigate the cause of this difference, we analyze the relevant physical parameters along the CME propagation paths. We consider two main forces that could account for this difference: the magnetic suppression force and the drag force. 
We define the top 10\% of the outermost points on the isosurface where $\mathrm{n/n_0}=3$ as the CME front (similar to \citealt{2020ApJ...895...47A}), and extract their positions at each time step. We then calculate the average local magnetic downward tension force $F_{\mathrm{tension},r}
= \frac{1}{\mu_0}\,(\mathbf{B}\cdot\nabla) B_r$
, wind density, and wind speed at these positions within the steady-state domain. The results are presented in Figure~\ref{fig:Fig.4}. Panel~(a) shows that the magnetic suppression near the equatorial region is weaker than at high latitudes. The stronger suppression in high-latitude regions can be attributed to the presence of stronger magnetic fields of opposite polarities in those areas. These results indicate that the large-scale magnetic field is unlikely to be responsible for the slower CME speeds at low latitudes. 
On the other hand, panels~(b) and~(c) show that the equatorial region is characterized by higher plasma density and lower wind speed, respectively, due to its location within a streamer region. Beyond 10 $R_\star$, the density of the equatorial streamer is about an order of magnitude higher than that of the high-latitude wind. For comparison, the density contrast in the solar case is more moderate, with the equatorial streamer density being roughly 3–4 times higher than that of the polar region wind \citep{1995GeoRL..22.3301P,1997JGR...10224151P}. Consequently, low-latitude CMEs experience stronger drag and are more difficult to accelerate. We therefore conclude that the background plasma density and wind speed are the primary factors responsible for the observed difference in CME dynamics between equatorial and high-latitude eruptions, rather than magnetic field suppression.

\begin{figure}
    \centering
    \plotone{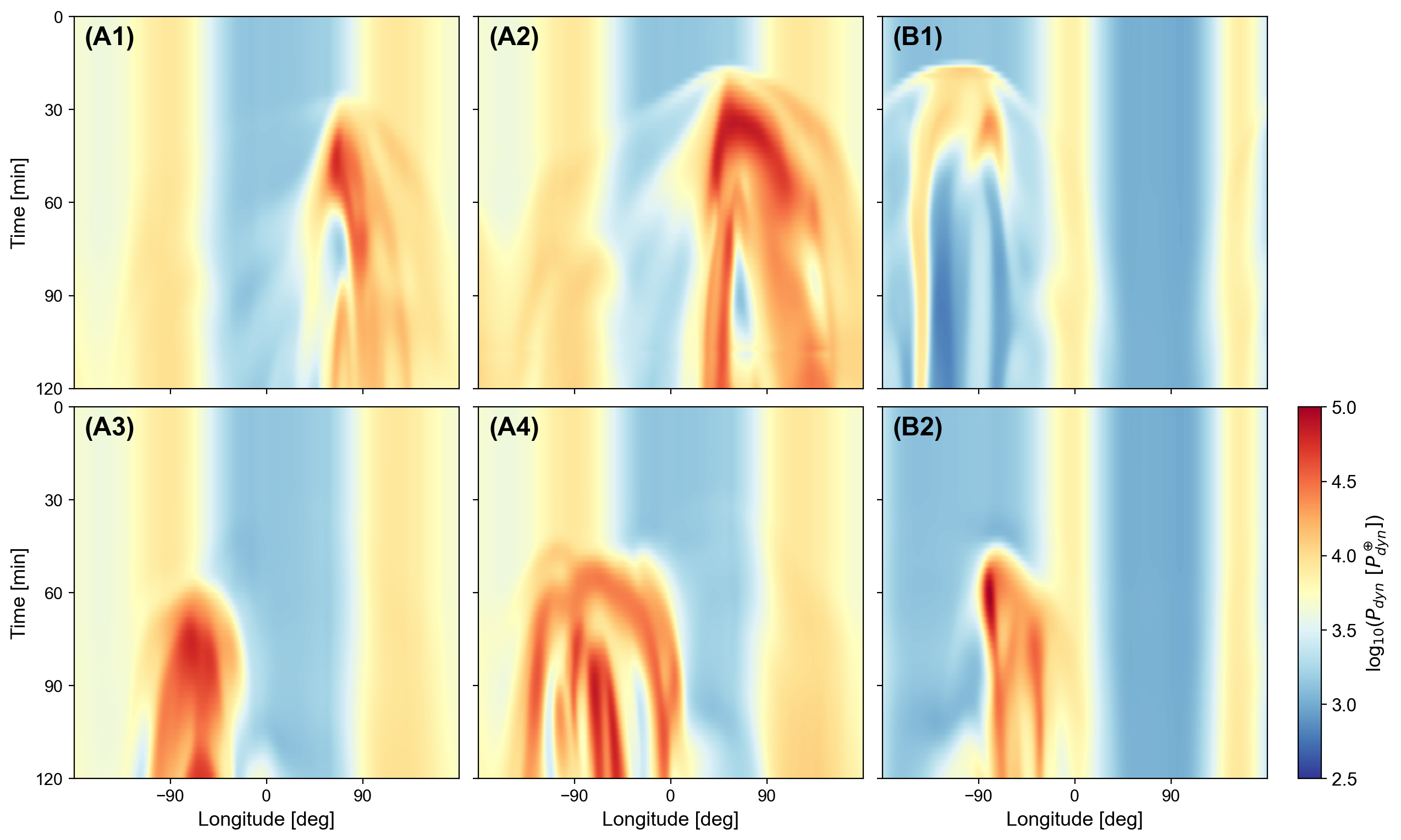}
  \caption{Time evolution of dynamic pressure on the equatorial orbit of the exoplanet. The colorbar is given in units of $P_{dyn}^{\oplus}$, the typical solar wind dynamic pressure at Earth.
    \label{fig:Fig.5}}
  \end{figure}

To assess the potential impact on exoplanets, we calculate the time-dependent dynamic pressure along the planetary orbit, defined as $P_{\mathrm{dyn}} = \rho v^2$. The results are presented in Figure~\ref{fig:Fig.5}.
For the high-latitude CMEs, the dynamic pressure on the equatorial plane peaks at $10^{4.78}$, $10^{4.87}$, and $10^{4.36}\ P_{\mathrm{dyn}}^{\oplus}$ for cases A1, A2, and B1, respectively. Interestingly, although the CME becomes more energetic from A1 to A2, the corresponding peak dynamic pressure does not increase significantly, but rather expands in spatial coverage (from $\sim 20\%$ to $\sim 40\%$ of the orbit). 
For the low-latitude CMEs, the dynamic pressure on the equatorial plane peaks at $10^{4.81}$, $10^{4.86}$, and $10^{5.0},P_{\mathrm{dyn}}^{\oplus}$ for cases A3, A4, and B2, respectively. The dynamic pressures associated with these equatorial CMEs are also relatively moderate. The spatial coverage of the perturbation is somewhat larger for A4 (about 50\%), but remains below 25\% for A3 and B2. It is worth noting that the dynamic pressure enhancement relative to the steady-state stellar wind in our simulation remains within two orders of magnitude. For comparison, in the M dwarf CME simulations by \citet{2022ApJ...928..147A}, the dynamic pressure enhancement can reach up to six orders of magnitude and with a much broader spatial extent (see their Figures~6 and~7). The discrepancy arises from differences in the large-scale magnetic field topology, which in our configuration causes fast CMEs to propagate toward higher latitudes and reduces their impact on the equatorial plane. In addition, we also calculated the resulting magnetic pressure and found that the dynamic pressure dominates, consistent with \citet{2025arXiv251020417H}.

\section{Discussion and Conclusion} \label{sec:kk}

In this work, we perform numerical simulations of stellar CMEs on a fully convective M dwarf using a rotation period of 30 days, which lies within the moderate-rotation regime. Based on the magnetic maps generated by an exploratory single-hemispheric dynamo model, we find that most PILs are located at high latitudes (40$^\circ$--55$^\circ$). In this configuration, CMEs are therefore expected to originate predominantly from high latitudes. Our results show that high-latitude CMEs primarily propagate toward higher latitudes, resulting in dynamic pressures near the equator that are both spatially limited and comparatively weak, even for more energetic CMEs (Case A2). Although some minor CMEs may originate from low latitudes, our simulations reveal that their propagation speeds are significantly reduced compared with the high-latitude ones. This behavior can be attributed to higher background plasma densities and weaker stellar winds in equatorial regions, which enhance drag and limit CME acceleration. As a result, the overall impact of these CMEs is minor, and the dynamic pressure experienced by equatorial exoplanets remains within two orders of magnitude above the steady-state stellar wind. This is substantially lower than the four to six orders of magnitude enhancement reported in previous M-dwarf CME simulations \citep{2022ApJ...928..147A}, owing to differences in the large-scale magnetic field topology, which lead to different evolution in the CME propagation. In conclusion, our findings suggest that if such single-hemisphere magnetic topologies exist in real M dwarfs and the escaping CME population is indeed dominated by eruptions from large-scale high-latitude PILs, equatorial exoplanets around these systems may reside in a relatively low-CME-impact environment, which might be favorable for long-term atmospheric retention and planetary habitability. 

The habitability of M-dwarf-hosting exoplanets remains a widely studied and debated topic.
Previous research has investigated various aspects of stellar space weather and its effects on planetary atmospheres, often suggesting that the intense space weather associated with M dwarfs may pose significant challenges to planetary habitability. Many studies have focused on the impacts of extreme ultraviolet (XUV) radiation (e.g., \citealt{2015Icar..250..357C,2017MNRAS.465L..74W,2023AstBu..78..588S,2025A&A...694A.310V}).
\citet{2023MNRAS.522.1411S} proposed the concept of a “UV habitable zone” and found that planets located within the traditional habitable zone of M dwarfs may lie outside this UV-habitable region. However, \citet{gao2026flare} showed that frequent stellar flares can enhance the near-UV flux and potentially extend the UV-habitable zone to overlap with the liquid-water habitable zone around low-mass stars. On the other hand, \citet{2025ApJ...985..100D} modeled atmospheric escape driven by stellar flares, showing that the atmospheres of exoplanets orbiting the active M dwarf AU~Mic could be depleted within a few billion years. In a broader context, these results can be interpreted within the framework of the ``cosmic shoreline'', which separates planets that can retain atmospheres from those that cannot as a function of stellar irradiation and planetary gravity \citep{2017ApJ...843..122Z,2025kiss.rept.....L}. Recent \textit{JWST} non-detections of atmospheres for several planets in M dwarf habitable zones may suggest that such planets lie on the atmospheric-loss side of this boundary (e.g., \citealt{2023Natur.618...39G,2023Natur.620..746Z}).

Other studies have examined the influence of stellar winds (e.g., \citealt{2013A&A...557A..67V,2014MNRAS.438.1162V,2016ApJ...833L...4G,2017ApJ...837L..26D,2020ApJ...902L...9A,2025A&A...701A.264W}). In our simulations, the steady-state stellar wind dynamic pressure experienced by exoplanets in the habitable zone reaches $\sim10^{3}$–$10^{4}\ P_{dyn}^{\oplus}$, consistent with previous findings (e.g., \citealt{2016ApJ...833L...4G,2022ApJ...928..147A}).
Such strong winds imply that planets would require significantly stronger intrinsic magnetic fields to sustain an Earth-sized magnetosphere. Nevertheless, recent observations indicate that some rocky planets orbiting M dwarfs can retain substantial atmospheres despite their extreme stellar environments \citep{2026arXiv260714326C}.
As for CMEs, M dwarfs have occasionally been observed to exhibit possible intense CME events (e.g., \citealt{1990A&A...238..249H,2022ApJ...933...92C,2025ApJ...978L..32L}). \citet{2022ApJ...928..147A} performed numerical simulations showing that CMEs from AU~Mic could generate dynamic pressures up to $10^{5}$ times stronger than those of extreme solar events at Earth.
\citet{2014ApJ...790...57C} found that M dwarf CMEs can cause substantial Joule heating in planetary atmospheres, while \citet{2025A&A...700A.225E} showed that CME-induced Joule heating on TRAPPIST-1e may exceed the average stellar XUV input by one to two orders of magnitude, potentially leading to severe atmospheric erosion. In contrast, our study provides a complementary perspective, suggesting that the stellar magnetic topology itself may play a critical role in shaping the space weather environment of exoplanets.
Specifically, stellar magnetic configurations dominated by high-latitude polarity inversion lines may naturally produce CMEs that propagate preferentially at high latitudes, thereby reducing their impact on planets in equatorial orbits. If such magnetic topologies occur in real stars, they could represent an additional factor influencing exoplanetary space weather environments and future habitability assessments.

Recent observations suggest that large flares on some fully convective M dwarfs preferentially occur at high latitudes. Using TESS photometry and geometric flare modeling, \citet{2021MNRAS.507.1723I} showed that several superflares on fully convective M dwarfs originated at latitudes above $\sim$55$^\circ$, indicating a non-random concentration of strong magnetic activity toward the stellar poles. Subsequent work by \citet{2024A&A...687A.138I} on one of these stars further suggested that the concentration of magnetic flux at high latitudes may reduce the observable coronal emission, potentially explaining its relatively low soft X-ray luminosity compared to typical low-mass stars in the X-ray saturation regime.
Similar patterns of high-latitude magnetic activity have also been reported for rapidly rotating solar-type stars. For example, Zeeman–Doppler imaging \citep{1997MNRAS.291....1D,1999MNRAS.302..437D,2007MNRAS.377.1488H} and X-ray observations \citep{2015ApJ...802...62D} of AB Dor reveal a pole-dominated magnetic field configuration. Moreover, a long-duration coronal dimming event detected on AB Dor \citep{2021NatAs...5..697V}, likely associated with a high-latitude CME, has been interpreted as originating from high latitudes, which could naturally account for its unusually long visibility. This scenario is further supported by numerical simulations of high-latitude CMEs in AB Dor \citep{2024MNRAS.533.1156S}. However, these cases involve much shorter rotation periods than our simulated star. Future dynamo simulations in the fast-rotation regime will be important for determining whether similar high-latitude magnetic field concentrations naturally arise under these conditions, thereby providing predictive constraints that can be directly tested by future high-resolution spectropolarimetric observations and time-domain flare and CME diagnostics.

Despite these findings, our results are subject to certain limitations. Most importantly, we emphasize that the adopted single-hemisphere dynamo topology should be regarded as an exploratory theoretical scenario rather than a representative magnetic configuration of fully convective M dwarfs. To date, similar magnetic topologies have not been conclusively confirmed observationally, nor widely reproduced in other global MHD dynamo simulations (e.g. \citealt{2016ApJ...833L..28Y,2021A&A...651A..66K}). Our results should therefore be interpreted primarily as an investigation of how such a hemisphere-asymmetric magnetic topology could influence CME dynamics and exoplanet space weather if it exists in real stars. 
Moreover, it should be noted that the adopted dynamo solution primarily represents a large-scale magnetic topology and does not include the small-scale magnetic structures that are commonly associated with local active-region PILs on the Sun. Therefore, small-scale active-region-related CMEs are not considered in this study. Nevertheless, the magnetic field surrounding the large-scale PILs can locally reach strengths of order kG, comparable to those of active regions. Because these PILs also extend over much larger spatial scales, they may store substantial amounts of magnetic free energy, potentially providing the magnetic environment required for the most energetic stellar eruptions \citep{2019ApJ...880...97L}. Previous numerical studies have shown that the strong global magnetic fields of active M dwarfs can substantially confine or even suppress CMEs originating from small-scale active regions \citep{2018ApJ...862...93A}. Consequently, the most energetic escaping CMEs may instead be associated with large-scale magnetic structures, which are the focus of the present study.
In addition, the benign space weather conditions inferred here apply only to exoplanets on equatorial orbits. However, observations indicate that exoplanetary orbits are not universally aligned with the stellar equatorial plane. Several systems exhibit significant spin-orbit misalignment, for instance, near-polar trajectories (e.g., GJ 436b, \citealt{2018Natur.553..477B,2022A&A...663A.160B}; HD 3167, \citealt{2019A&A...631A..28D}). Additionally, the population of moderately rotating fully convective M dwarfs may be comparatively sparse relative to the rapidly and slowly rotating populations \citep{2018AJ....156..217N,2019A&A...621A.126D}. Nevertheless, ongoing efforts in high-resolution spectroscopy and polarimetry, such as those enabled by SPIRou and NIRPS \citep[e.g.,][]{2018MNRAS.475.1960F,2023MNRAS.525.2015D,2025A&A...700A..11S}, are expected to significantly expand the sample of magnetically-characterized low-mass stars and their planetary systems in the near future. Furthermore, the advent of next-generation extremely large telescopes (e.g., ELT, GMT, and TMT) and their advanced instrumentation will provide unprecedented sensitivity to probe the magnetic properties of moderately-rotating fully convective M dwarfs, potentially revealing many more examples of such systems and improving our understanding of stellar magnetism across the low-mass regime. Complementary radio observations with facilities such as FAST and LOFAR can further diagnose stellar coronal magnetic fields and particle acceleration processes through coherent radio emission \citep{2023ApJ...953...65Z,2025SciA...11.6116Z,2024NatAs...8.1359C,2025Natur.647..603C}, offering an independent probe of stellar magnetic topologies.
  
\acknowledgments
We thank the anonymous referee for the helpful comments and suggestions. This work is supported by the National Natural Science Foundation of China (12425301 \& 425B2024), and the Specialized Research Fund for State Key Laboratory of Solar Activity and Space Weather. H.T. is also supported by the New Cornerstone Science Foundation through the Xplorer Prize. K. P. acknowledges that this work was co-funded\footnote{Views and opinions expressed are however those of the author(s) only and do not necessarily reflect those of the European Union or the European Research Council. Neither the European Union nor the granting authority can be held responsible for them.} by the European Union (ERC-CoG, Evaporator, No. 101170037). This work was carried out using the SWMF and BATS-R-US tools developed at the University of Michigan’s Center for Space Environment Modeling. We also thank the technical support of ``National Large Scientific and Technological Infrastructure Earth System Numerical Simulation Facility"\footnote{https://cstr.cn/31134.02.EL}.

\bibliographystyle{aasjournal}
\bibliography{ref}

\end{document}